\documentclass[conference]{IEEEtran}
\IEEEoverridecommandlockouts

\usepackage[utf8]{inputenc}
\usepackage{hyperref}
\usepackage{xurl}
\usepackage{orcidlink}
\usepackage{subcaption}
\usepackage{eso-pic}
\usepackage[numbers,sort&compress,comma,square]{natbib}

\newcommand{\figref}[1]{Figure~\ref{#1}}
\newcommand{\coo}{\ensuremath{\mathrm{CO_2}}}

\newcommand{\hreffinternal}[3]{\href{#1}{\textcolor{#3}{#2}}}
\newcommand{\hreff}[2]{\hreffinternal{#1}{#2}{blue}}

\newcommand{\placetextbox}[3]{% \placetextbox{<horizontal pos>}{<vertical pos>}{<stuff>}
  \setbox0=\hbox{#3}% Put <stuff> in a box
  \AddToShipoutPictureFG*{% Add <stuff> to current page foreground
    \put(\LenToUnit{#1\paperwidth},\LenToUnit{#2\paperheight}){\vtop{{\null}\makebox[0pt][c]{#3}}}%
  }%
}%

\begin{document}

\placetextbox{0.5}{0.99}{\large\colorbox{gray!10}{\textcolor{red}{\textbf{Author pre-print.}}}}%

\placetextbox{0.5}{0.97}{\large\colorbox{gray!10}{\textcolor{red}{\textbf{Article accepted for the New Ideas and Emerging Results (NIER) track of \hreff{https://conf.researchr.org/home/icse-2025}{ICSE 2025}.}}}}%

\placetextbox{0.5}{0.05}{\colorbox{gray!10}{\textcolor{red}{\textbf{Author pre-print.}}}}%

\title{SusDevOps: Promoting Sustainability to a First Principle in Software Delivery}

\author{\IEEEauthorblockN{Istvan David}
\thanks{We acknowledge the support of the Natural Sciences and Engineering Research Council of Canada (NSERC), DGECR-2024-00293.}
\IEEEauthorblockA{\textit{Sustainable Systems and Methods Lab, McMaster University}\\
\textit{McMaster Centre for Software Certification (McSCert)}\\
Hamilton, Canada -- istvan.david@mcmaster.ca}}

\maketitle

\begin{abstract}
  Sustainability is becoming a key property of modern software systems. While there is a substantial and growing body of knowledge on engineering sustainable software, end-to-end frameworks that situate sustainability-related activities within the software delivery lifecycle are missing. In this article, we propose the SusDevOps framework that promotes sustainability to a first principle within a DevOps context. We demonstrate the lifecycle phases and techniques of SusDevOps through the case of a software development startup company.
\end{abstract}

\begin{IEEEkeywords}
DevOps,
software delivery,
sustainability
\end{IEEEkeywords}

\section{Introduction}\label{sec:intro}

Estimates show that the Information and Communications Technology (ICT) sector currently contributes to about 2–4\% of global CO2 emissions, and this number is projected to increase to about 14\% by 2040~\cite{belkhir2018assessing}. To follow suit with the rest of the global economy, the ICT sector should—directly or indirectly—decrease its CO2 emissions by 42\% by 2030, by 72\% by 2040, and by 91\% by 2050~\cite{ituict}.
These numbers must concern software practitioners for a number of reasons.

First, the nature of user requirements is changing.
While sustain-ability-related user requirements are not quite mainstream currently, sustainability is shaping up to become the non-functional requirement of the 21st century~\cite{penzenstadler2014safety}. The expectation that users and organizations will not only reward but demand efforts toward sustainability has been identified as the top ``global megatrend'' by the International Council on Systems Engineering~\cite{incose2022vision}, necessitating a radically new approach to the engineering of software and software-intensive systems.
Second, even if a company embraces the idea of developing sustainable software, the lack of software delivery frameworks that can accommodate sustainability goals quickly becomes a show-stopper. Relating software functionality to sustainability goals along a software development lifecycle is not a trivial endeavor. When should sustainability requirements be addressed? How should they be prioritized? What techniques and tools can support systematic decision-making?

There is a substantial and rapidly growing body of knowledge on engineering sustainable software, with high-quality and actionable methods and techniques~\cite{penzenstadler2018software,lago2015framing}. This body of knowledge is ready to be put to use. To achieve this, we need end-to-end frameworks that promote sustainability to a first principle, instead of treating it as a quality metric~\cite{david2024circular}.

In this paper, we propose such a framework, called SusDevOps. As the name suggests, SusDevOps builds on the established software development and operations practices of DevOps [5] and aligns sustainability practices (``Sus'') with them. The framework defines a holistic software development lifecycle model to weave these concerns into a coherent unit.

We aim to provide useful information to three groups of practitioners. Software developers, so that they can anticipate the skills required in a sustainability-first era of software delivery. Architects, product owners, and delivery process owners, so that they can understand how to extend delivery processes to incorporate activities in support of sustainability ambitions. And decision-makers, who might have already identified the need for sustainability and are looking to revamp their operations for sustainability-first software practices.

\iffalse
\id{Merge this in here:}
Sustainability is usually characterized by three complementary dimensions, first defined in the Brundtland report~\cite{brundtland1987our}. Economic sustainability is concerned with the financial viability of a software product; environmental sustainability focuses on reduced ecological impact, such as energy consumption; and social sustainability promotes the elevated utility of software for humans. \citet{penzenstadler2013generic} define a fourth dimension, technical sustainability, focusing on the prolonged service time of software systems, chiefly supported by proper evolution methods. All of these four dimensions must be considered to achieve sustainability in software systems.
Software engineers and practitioners are typically focused on technical sustainability, i.e., evolvability and maintainability, as these concerns are closest to software code and decisions can be made in the scope of a software engineer's scope of authority~\cite{david2023towards}. However, there is an emerging awareness of environmental sustainability, such as the energy consumption of software, as resource-intensive methods, such as blockchain and large-scale machine learning become parts of modern software systems. Tools and methods in support of energy assessment of software~\cite{kalaitzoglou201408practical} and energy-aware design are increasingly easier to incorporate into everyday software engineering practices.
\fi
\section{Background on DevOps}

DevOps~\cite{ebert2016devops} is the collection of values, principles, practices, and tools that narrows the gap between developing (``Dev'') and operating (``Ops'') software systems. By promoting rapid iterative and incremental practices, DevOps increases the efficiency of delivery. While terminology is not standardized, the main stages of DevOps are well-understood by practitioners. These stages are the following.
\textit{PLAN} -- definition of process metrics for management and the definition of technical requirements for the software.
\textit{CREATE (or CODE)} -- development of the software system.
\textit{VERIFY (or TEST)} -- quality assurance, testing, validation, and verification.
\textit{PACKAGE} -- package configuration, release staging, and release approvals.
\textit{RELEASE} -- moving the software into production, including coordination, fallbacks, and recovery.
\textit{CONFIGURE} -- preparation and provisioning of the hosting infrastructure.
\textit{MONITOR} -- keeping track of (non-)functional metrics, e.g., performance, response time, and end-user experience.
Feedback from monitoring is used in the planning of the next deliverable.

DevOps and its various augmentations are widely adopted. Such approaches include
BizDevOps~\cite{gruhn2015bizdevops}, connecting business operations (``Biz'') to software delivery;
DevSecOps~\cite{alonso2023embracing}, focusing on security (``Sec'');
and recently, MLOps~\cite{kreuzberger2023machine}, that applies DevOps principles to machine learning.
%CloudOps~\cite{alonso2022cloudops}, DataOps~\cite{munappy2020ad}.

Closest to our proposal is GreenOps~\cite{greenops}, which integrates technologies and business practices to maximize efficiency in the cloud while reducing environmental impact. However, GreenOps does not define a lifecycle model.
\section{The SusDevOps Framework  -- A Proposal}

SusDevOps is a software delivery framework that treats sustainability as a first principle of software. As the name suggests, the framework builds on DevOps practices of integrated software development (``Dev'') and software operation (``Ops'') and extends them with the practice of sustainability (``Sus''). To keep software delivery agile despite the distinguished role of sustainability, SusDevOps defines a lifecycle model (Figure 1) that integrates sustainability-related activities with the software development and delivery activities of traditional DevOps.

In SusDevOps, sustainability is not a mere metric but a goal that drives product design and delivery. This is an important improvement over current software delivery frameworks. For example, in traditional DevOps, sustainability is approached as a quality metric of the software~\cite{lago2015framing}; and while BizDevOps~\cite{gruhn2015bizdevops} improves over this by treating sustainability as a business goal rather than a technical quality metric, it fails to connect sustainability to software requirements.

\begin{figure}[h]
    %\vspace{-0.75em}
    \centering
    \includegraphics[trim={6.75cm 2.5cm 11.25cm 1.75cm},clip,width=0.7\linewidth]{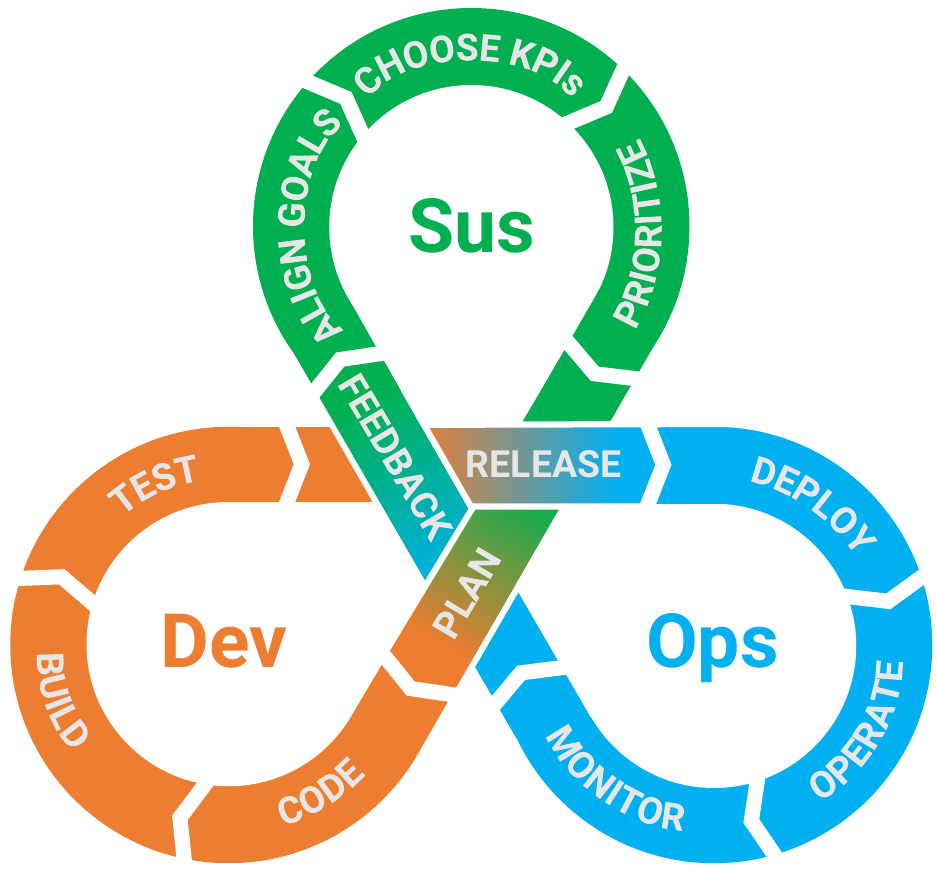}
    \caption{The SusDevOps framework}
    \label{fig:susdevops}
    %\vspace{-0.75em}
\end{figure}

In the following, we elaborate on the distinguishing elements of SusDevOps, i.e., the activities of the ``Sus'' phase.
%The key sustainability-related activities of SusDevOps are the following.

\paragraph*{\textbf{Align goals}} First, sustainability goals are formulated based on the feedback from operations and are aligned with the goals of the product. Sustainability goals are more often contradicting than technical requirements and the resolution of these contradictions requires a dedicated activity. This requires a holistic, system-level view, as sustainability goals touch upon technical aspects and business operations as well. In the illustrative case, for example, environmental sustainability is articulated as a key expectation from the users. Aligning it with the technical parameters of the software is not trivial and requires understanding the impact of environmentally-friendly software design on functional and extra-functional properties.

\paragraph*{\textbf{Choose KPIs}} After sustainability goals have been articulated, they need to be properly underpinned by measurable key performance indicators (KPIs). KPIs are indicators focusing on the aspects of software systems that are most critical for the success of the company~\cite{parmenter2015key}. Establishing the right KPIs is challenging, but solutions exist. For example, \citet{fatima2024providing} provide practitioners with a template-based tool.
%In the case, \coo{} emission has been chosen as a heuristic for environmental sustainability. This can be now tied to the energy consumption of the software.

\paragraph*{\textbf{Prioritize goals}} Finally, priorities among goals are set. Typically, companies will face trade-off questions between sustainability and revenue, performance, etc. Dedicating a specific step to resolve these questions is necessitated by the potential involvement of higher-level decision-makers.

\section{Illustrative case}

We use the example of a software development startup company to illustrate the usage of the SusDevOps framework in a typical business-to-consumer (B2C) setting.
The company develops a software product of which the key feature is high-precision simulation driven by pre-trained AI models. User satisfaction is mainly influenced by the precision of the product. Being a startup, the company uses the majority of their revenue to regularly improve their computing capacity to provide better precision through better-trained AI models.
After the user base grew large enough, the company noticed that user satisfaction is also influenced by the perceived environmental sustainability of the software, such as energy efficiency. Given the B2C model, revenue might be significantly impacted by emerging user needs, leading to less liquidity to improve the product.

The main challenge in this case is to stay agile in the delivery process, that is, to maintain the right velocity while reacting to sustainability needs as they emerge. These needs have to be assessed systematically, trade-offs between technical and sustainability goals need to be identified, those trade-offs have to be mapped onto the functionality of the software through requirements, etc. Traditional software engineering delivery processes might struggle to handle this challenge efficiently due to the convoluted and vague notion of sustainability.

These challenges can be addressed through the activities along the lifecycle defined by the SusDevOps framework.
%In the following, we briefly demonstrate these activities.

\subsection{Monitor (Ops) and Feedback (Ops-to-Sus)}
Monitoring is a DevOps activity that helps keep track of functional and non-functional metrics. In SusDevOps, this activity also includes keeping track of sustainability properties (e.g., the energy consumption of software), and the satisfaction of end-users. In anticipation of emerging sustainability needs of users, companies can run market studies and use the feedback functionality of applications to gather valuable leads. In our case, a market study finds that users value perceived sustainability highly.
This information is fed back to the development team to support the planning of the next release.

\subsection{Align goals (Sus)}
In the first sustainability activity, goals are formulated based on the feedback and are aligned with the goals of the product. Causal loop diagrams (CLD) are an appropriate formalism to capture system dynamics. A practical and accessible overview of using CLDs in software engineering has been provided by~\cite{penzenstadler2018software}.
As the name suggests, CLDs aim to visualize the causal relationships between system variables. System variables might influence each other through positive or negative causal loops, meaning that two variables change in the same or opposite direction, respectively.

\figref{fig:cld-0} shows the original system before feedback. Precision is a key influencing factor of User satisfaction: the higher the precision, the higher the user satisfaction. This causal relationship is shown by the link between the variables denoted by a + sign. By the same logic, higher User satisfaction leads to higher Revenue which, in turn, allows the company to further increase the Precision of the software. This leads to a reinforcing loop (denoted by “R”) of Technical performance. Reinforcing loops are associated with exponential increases and decreases—in this case, with a rapid increase in revenue. Of course, hardware limitations and market saturation keep the system in check, and variables plateau at some point.

Aligning the new goal means adding \textit{Perceived sustainability} to the system to form loops in later steps. \textit{Perceived sustainability} positively influences \textit{User satisfaction}. To form loops and to further detail the diagram, we need more details.

\subsection{Choose KPIs (Sus)}
\begin{figure}[h]
    \centering
    \begin{subfigure}{\linewidth}
        \centering
        \includegraphics[width=0.6\linewidth]{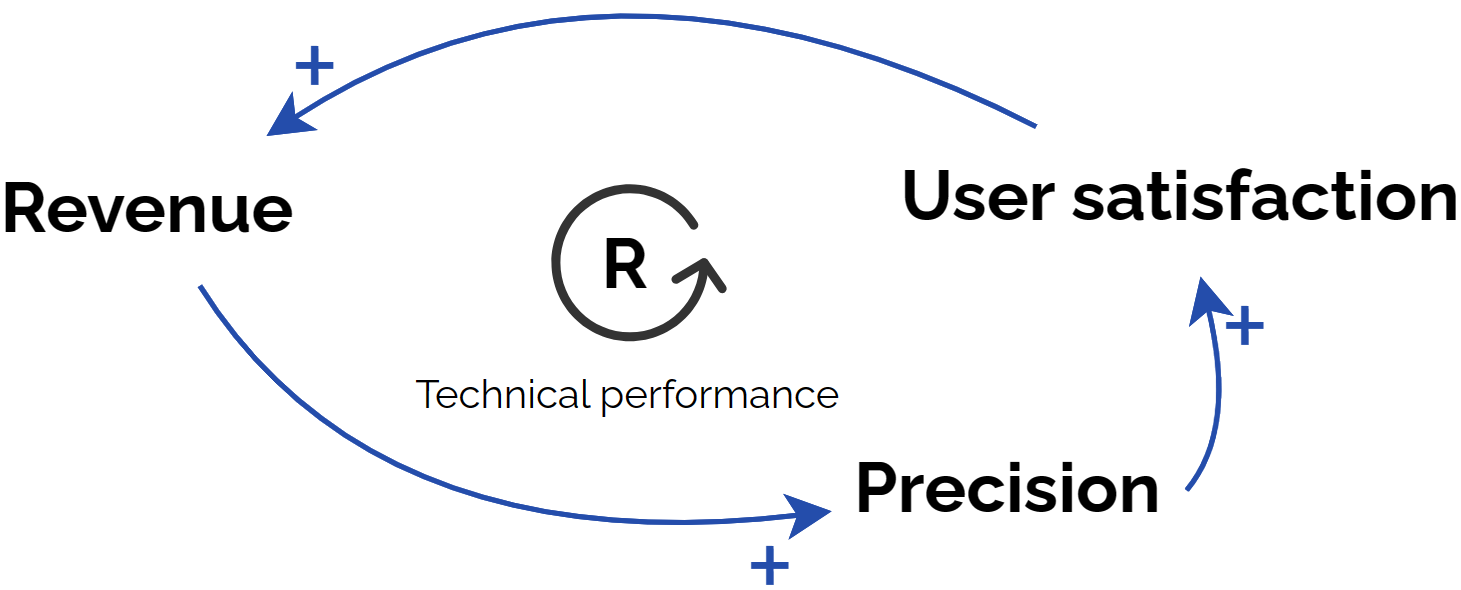}
        \caption{Original understanding before feedback}
        \label{fig:cld-0}
    \end{subfigure}
    \begin{subfigure}{\linewidth}
        \centering
        \includegraphics[width=\linewidth]{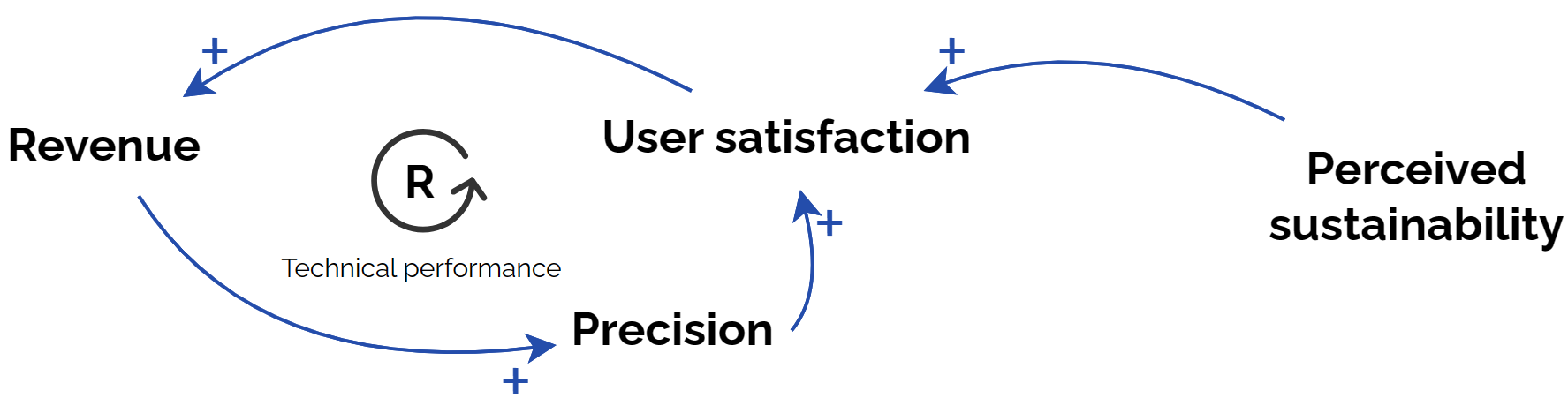}
        \caption{Updated understanding after feedback}
        \label{fig:cld-1}
    \end{subfigure}
    \begin{subfigure}{\linewidth}
        \centering
        \includegraphics[width=\linewidth]{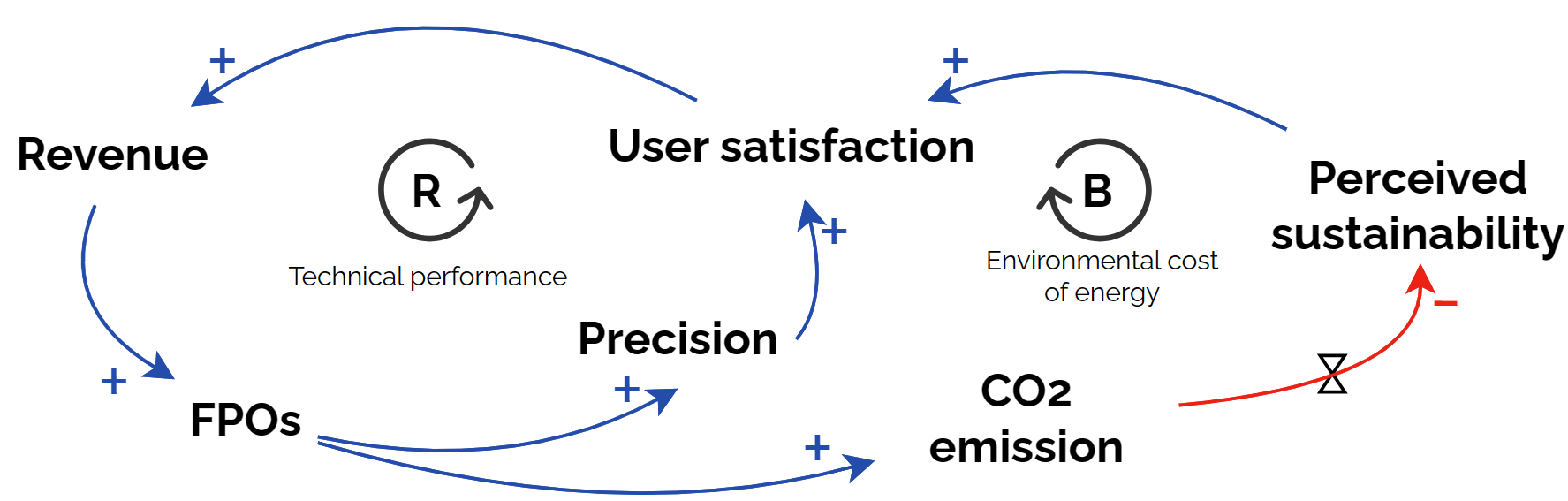}
        \caption{New KPIs (FPO, \textbf{\coo{}} emission) and new links form a new loop}
        \label{fig:cld-2}
    \end{subfigure}
    \caption{Evolving understanding of sustainability factors}
    \label{fig:case}
\end{figure}

After the new goal has been articulated, it has to be properly elaborated by choosing the right KPIs. KPIs are indicators focusing on the aspects of software systems that are most critical for the success of the company~\cite{parmenter2015key}.
In our case, \figref{fig:cld-1} is extended by two KPIs, shown in \figref{fig:cld-2}.

To approximate environmental sustainability, the \coo{} emission of computation is used as a KPI. \coo{} emission impacts perceived sustainability negatively, hence, a negative relationship is drawn in the causal loop diagram. To approximate precision, the number of floating-point operations (FPO) can be used as a KPI. Common choices are the cumulative energy consumption and cumulative power consumption of computation as well, but the relationship between energy, power, and run time is ambiguous. Power (SI-unit: watts) is the amount of work (SI-unit: joules) over time. Very simplistically, we say power = work / time. Conversely, energy = power × time. Reducing computational resources results in lower power consumption, but prolongs computation time; thus, due to the latter equation, the change in energy is unclear. Adding more resources to reduce computation time will result in shorter computation time but higher power consumption and again, the change in energy is unclear. FPO calculates the amount of work needed for computation directly and is thus tied to the amount of energy consumed~\cite{schwartz2020green}, and in the long term, to \coo{} emissions required to produce this energy. This now allows for establishing two new links.

First, the link between Revenue and Precision is replaced by a link between Revenue to FPO, and FPO to Precision. The effect of influences does not change, as every link remains a positive causal link: with higher revenue, the company can afford more FPOs, leading to higher precision, and increasing user satisfaction which, in turn, leads to increased revenue—forming a reinforcement loop. Second, a new link is placed between FPO and \coo{} emission. The link is a positive causal link: the higher the FPO, the higher the \coo{} emissions are. This now forms a new balancing loop (denoted by “B”). That is, with the increasing \coo{} emission, Perceived sustainability decreases, eventually leading to lower Revenue, lower number of FPOs, and lower \coo{} emissions. Eventually, the loop will find a balance among the variables and stabilize.

\subsection{Prioritize (Sus)}

In the final activity of the sustainability practice of SusDevOps, priorities among goals are set.
The company faces an important question: what is the right trade-off between precision and perceived sustainability? This question emerged solely because the previous activity identified a balancing loop in the system. From Figure 2c we know that both precision and \coo{} emissions can be expressed in relation to the number of FPOs a computation requires to achieve its goals. The challenge is that parameter sensitivity between precision and \coo{} emission is not known unless empirical data is collected first. In alternative terms, it is not known how much precision and \coo{} emission will decrease by a unit of decrease of FPO. Evidence from the AI domain~\cite{schwartz2020green} suggests that by just sacrificing 0.5\% of precision, FPO can decrease by as much as 30-35\%. It is still not entirely clear how this decrease in FPO will impact \coo{} emissions, but surely, it is a step in the right direction. Further experiments can shed light on this relationship in the context of the specific software product.

Factors that commonly influence prioritization are user needs, business goals, corporate values, and in some cases, laws and regulations. Specifically, corporate values influence the leverage point a company chooses to influence the ecosystem their software product is a part of. For example, following the leverage point clusters of~\cite{penzenstadler2018software}, a company might not even wait for emerging user needs, but take a proactive stance and choose the highest leverage points to change the intent of the system and stakeholders. This could be achieved by raising sustainability awareness, for example, through gamification that rewards more energy-efficient usage of the software.

\subsection{Plan (Sus-to-Dev)}
To conclude the sustainability phase of the SusDevOps process, plans are formulated based on aligned, elaborated, and prioritized goals. In a typical software engineering setting, this mainly means formulating requirements or change requests for the software~\cite{schwartz2020green}. This activity also includes formulating engagement plans with users to advocate change and bring users on board with the new, more sustainable product.

Eventually, the cycle is completed with the traditional DevOps activities of software development and operation.

\section{Discussion}

%In this article, we proposed a novel, sustainability-first software development and delivery framework, SusDevOps.

SusDevOps is a novel, sustainability-first software delivery framework that defines an integrated lifecycle model based on well-established DevOps principles. Through the case of a software development startup company, we demonstrated the utility of the proposed framework, its alignment with standard software development and operation processes, and recommended techniques and tools to cover the various sustainability-related stages of the lifecycle.
Of course, SusDevOps is no silver bullet and might not serve as a blueprint for every software project. However, it provides a reasonably general framework to organize team effort in delivering sustainable software.
Adopters should maintain a pragmatic standpoint and implement SusDevOps through techniques and tools that fit their goals, skills, and organizational capabilities.

%Employing SusDevOps has its distinct benefits and challenges. We now review the most important ones.

\subsection{Benefits of using SusDevOps}

Some of the benefits of SusDevOps are the following.

\subsubsection{End-to-end understanding} The framework aligns elementary software delivery activities with sustainability activities. This allows stakeholders along the software delivery process (including IT departments) to understand when and how to deal with sustainability, what input to gather for sustainability-related decision-making, and which sustainability goals to translate to software requirements.

\subsubsection{Actionability} The framework defines activities for the sustainability practice, and for each activity, it recommends techniques and tools. This lowers the barriers to adopting SusDevOps and simplifies aligning it with business processes.

\subsubsection{Agility} The framework promotes systematic yet rapid sustain-ability-related decision-making in the iterative-incremental DevOps context. This allows for reacting to change—which is very much pronounced in end-user sustainability requirements—more efficiently.

\phantom{}

The sequential alignment of the Sus and Dev practices might challenge the agility of SusDevOps. The guiding principle of SusDevOps is that requirements are first investigated through the filter of sustainability before reflecting on their technical aspects. These ideas have been recognized in sustainable software engineering, especially in the context of requirements elicitation, e.g., by \citet{becker2016requirements}. Promoting sustainability to such a prominent position, and treating it as a principle practice along with development and operations, fosters a culture shift in which new techniques and tools in support of delivering sustainable software emerge naturally. Iterations between the Sus and Dev practices before releasing the product might be still introduced if needed, similar to the iterations in the Dev phase in traditional DevOps before release.

\subsection{Challenges in adopting SusDevOps}

Here are some of the key challenges practitioners should anticipate when adopting SusDevOps.

\subsubsection{Proper Monitoring and Feedback mechanisms should be identified}
The lack of mechanisms to gauge end-users’ sustainability needs hinders the agility of the delivery process and slow it down. In B2B settings, this challenge can be addressed by working closely with clients. B2C settings, however, necessitate innovative ways to engage with users.

\subsubsection{New competencies need to be developed}
While sustainability expertise might be brought in as a service, it is more sustainable for a company to develop internal skills. This entails expanding the skillsets of business analysts, product owners, and software engineers with the foundational concepts, techniques, and tools that help along the sustainability practice of SusDevOps. In the illustrative case, each sustainability-related step was executed by these internal professionals.

\subsubsection{Stakeholder buy-in is paramount}
As sustainability transcends the software’s purpose~\cite{becker2016requirements}, it requires active support from all stakeholders in a company. However, the stratified nature of sustainability (i.e., having different meanings at different levels of organizational hierarchy) challenges such efforts.
Participatory and collaborative modeling are key enablers in the design for sustainability with non-technical stakeholders on board~\cite{manellanga2024participatory}.
To simplify the challenge of joint decision-making in the convoluted problem of sustainability, \citet{becker2016requirements} recommend focusing on stakeholders with influence.

\subsubsection{Knowledge management is key}
Sustainability goals are inherently interdisciplinary and multisystemic: their sound interpretation is possible only by investigating them from multiple angles. Even the simple illustrative case in this article touched upon business goals, technical goals, and environmental goals. Thus, facilitating processes to continuously maintain the map of sustainability goals is key to having the proper understanding of the sustainability of the developed software within its extended socio-technical context.

\iffalse
\subsubsection*{Watch out for accidental greenwashing.}
Greenwashing is the deceptive strategy of magnifying a company’s sustainability efforts. Greenwashing can be accidental too, for example, by focusing on a single environmental attribute while ignoring others. Understanding sustainability goals and their alignment with the overall product strategy, and subsequently choosing the right KPIs is a key moment in making sound sustainability decisions and by that, avoiding accidental greenwashing. In a recent article, \citet{fatima2023goals} provide actionable pointers and a template-based tool for developing KPIs. It is important that adopters of SusDevOps embrace continual improvement. Especially in convoluted problems, such as the sustainability of software, the best KPIs can be developed over time.
\fi

\iffalse
\input{tables/table1}

\footnotetext[1]{https://www.visual-paradigm.com/}
\footnotetext[2]{https://miro.com/}
\footnotetext[3]{https://vensim.com/}
\fi
\section{Future Plans}

To foster the adoption of SusDevOps, we focus on three distinct future directions.
(1) Tools should be proposed through which stakeholders can participate in the SusDevOps process. Such tools range from technical ones (e.g., modeling tools~\cite{bork2024role,lago2024sustainability}
%for causal loop diagrams)
to productivity and management tools (e.g., KPI design templates~\cite{fatima2024providing}).
(2) Organizational enablers at potential adopters should be investigated, especially under the broader umbrella of Twin Transition~\cite{david2024sustainable}. We are currently running a survey on this topic.
(3) Finally, we will rigorously evaluate the approach in collaboration with industry partners.

\bibliographystyle{IEEEtranN}
\bibliography{bib/references}

\end{document}